\documentstyle[12pt]{amsart}
\title{On Static $n$-body Configurations in Relativity}

\newtheorem{thm}{Theorem}[section]

\newtheorem{prop}[thm]{Proposition}

\theoremstyle{definition}

\numberwithin{equation}{section}

\renewcommand{\a}{\alpha}

\renewcommand{\d}{\delta}

\newcommand{\e}{\varepsilon}
\newcommand{\g}{\gamma}

\newcommand{\p}{\partial}

\newcommand{\s}{\sigma}

\def\Pb{\ifmmode{\Bbb P}\else{$\Bbb P$}\fi}
\def\Z{\ifmmode{\Bbb Z}\else{$\Bbb Z$}\fi}
\def\C{\ifmmode{\Bbb C}\else{$\Bbb C$}\fi}
\def\R{\ifmmode{\Bbb R}\else{$\Bbb R$}\fi}
\def\S{\ifmmode{S^2}\else{$S^2$}\fi}

\def\S{\cal S}

\title[Static Configurations]{On static $n$-body configurations in relativity}
\author {Robert Beig and Richard M. Schoen}
\address{R. Beig\\ Faculty of Physics\\ Gravity Group\\ University of Vienna\\ Vienna}
\address{R. Schoen\\ Department of Mathematics\\ Stanford University\\ Stanford, CA 94305}
\thanks{This work was the result of a collaboration which occurred at the Mittag-Leffler Institute in the Fall of 2008. The authors thank the institute for its support. The first author was also partially supported by  Fonds zur F\"orderung der wissenschaftlichen Forschung, project P20414-N16. The second author was partially supported by the Clay Mathematics Institute.}

\begin{document}
\maketitle

\setcounter{secnumdepth}{1}

\begin{abstract}
The static $n$-body problem of General Relativity states that
there are, under a reasonable energy condition, no static $n$-body
configurations for $n > 1$, provided the configuration of the bodies
satisfies a suitable separation condition. In this paper we solve this problem
in the case that there exists a closed, noncompact, totally geodesic surface disjoint
from the bodies. This covers the situation where the configuration has a reflection
symmetry across a noncompact surface disjoint from the bodies.
\end{abstract}

\setcounter{section}{0}

\section{\bf{Introduction and background}}
A classical result in Newtonian gravity is that there can be no static $n$-body configuration for which the bodies are separated by a plane disjoint from the bodies. On the other hand one can concoct static $2$-body configurations in Newtonian theory \cite{cel} with both bodies being contractible and one body sufficiently non-convex so that the convex hulls of the bodies intersect.  Analogous configurations exist for relativistic bodies (work in progress by
L. Andersson, the first author, and B. G. Schmidt). For $n > 1$ and assuming a suitable energy condition, it is reasonable to conjecture a relativistic analogue of the Newtonian result stated above; that is, $n$-body static configurations should be impossible provided some separation condition for the bodies is satisfied. The work \cite{mueller} has some results on the static $n$-body conjecture, but no theorem under easily stated conditions. In the present paper we show (see Theorem 2.2) that an asymptotically flat triple $(M,V,g)$ with nonnegative scalar curvature which is static vacuum outside a compact set and in a neighborhood of a closed, embedded, noncompact, totally geodesic surface is trivial. This solves the static $n$-body problem in the case that the configuration has a reflection symmetry across a noncompact surface which is disjoint from the matter regions (see Theorem 2.3).

Recall that static spacetimes are $4$-manifolds with a metric of Lorentz signature which have a Killing
vector field which is complete, everywhere timelike, and hypersurface orthogonal. General Relativity studies such spacetimes subject to the Einstein equations $G_{\mu\nu} = 8 \pi G T_{\mu\nu}$ (see \cite{wald}). Such solutions describe the gravitational fields of time independent, non-rotating sources.
Static spacetimes can be written as warped products ${\mathbb R} \times M$ with metric $ds^2$ of the form
\begin{equation}
ds^2 = - V^2 (x)\, dt^2 + g_{ij}(x)\, dx^i dx^j
\end{equation}
with $V$ a positive function and $g$ a Riemannian metric on the $3$-manifold $M$. The Einstein equations then take the form
\begin{equation}
\Delta V = 4 \pi G V (\rho + \tau) \label{DelV}
\end{equation}
and
\begin{equation}\label{Rij}
V R_{ij} - D_i D_j V = 4 \pi G V \left[(\rho - \tau)\, g_{ij} + 2 \tau_{ij}\right]\,,
\end{equation}
where $\rho$ and $\tau_{ij} = \tau_{(ij)}$ are respectively the
energy density and the stress tensor in the rest system of the
matter and $\tau = \tau_i{}^i$ is the trace. We are interested in
solutions to these equations corresponding to $n$ isolated bodies.
By this we mean the following: First the 3-manifold $(M,g)$ is
asymptotically flat with $V$ tending to 1 at infinity.
Secondly
we assume that the support of the matter fields $\rho,\,\tau_{ij}$
is contained in $n$ disjoint compact connected sets
$\overline{\Omega_r}$, with $\Omega_r$ open with smooth boundary
$\partial \Omega_r$ for $r=1,\ldots, n$. Finally we assume that all
fields are sufficiently smooth (even analytic) except across
$\partial \Omega_r$ where $\rho,\,\tau_{ij}$ and the normal
components of $\partial^2g_{ij},\,\partial^2 V$ will in general have
jump discontinuities. We require also that $g$ and $V$ be $C^1$
across the boundaries. Let us remark that taking the trace of (1.3)
and using (1.2) we recover the time symmetric initial value
constraint
\begin{equation}
R = 16 \pi G \rho
\end{equation}
and taking a divergence of (1.3), using (1.2) and the contracted Bianchi identity,
we find that
\begin{equation}
D_j (V \tau_i{}^j) + \rho\, D_i V = 0\,,
\end{equation}
which plays the role of equilibrium condition for the matter variables.
In order for this condition to hold distributionally across the boundaries we require the additional boundary condition
\begin{equation}
\tau_i{}^j n_j|_{\partial \Omega_r} = 0\,
\end{equation}
that is, the stress should have zero normal components to the
boundary of the bodies. In many models of continuum mechanics the
stress tensor is a functional of a collection of matter fields and
their first derivatives, which renders equation (1.5) a quasilinear
second order PDE with Neumann-type boundary conditions (1.6). For
perfect fluids one has that $\tau_{ij} = p\, g_{ij}$ with $p > 0$ in
$\Omega_r$ and $\rho$ a given positive non-decreasing function of
$p$ in $\mathbb{R}^+$. There are different energy conditions which
one might impose on the matter variables (see \cite{he}), the
weakest one being that $\rho \geq 0$, which is sufficient for the
positive mass theorem \cite{sy} to be valid. Finally one might
mention here the case of black holes, in which the regions $\cup_r
\Omega_r$ are missing, but instead at the boundaries $V|_{\partial
\Omega_r} = 0$ with $\partial \Omega_r$ being totally geodesic
surfaces.

Historically, the 'no-body situation', i.e. $n=0$, implies that $(M,V,g)$ is trivial (Minkowski) in the sense that $V = 1$ and $(M,g)$ is flat ${\mathbb R}^3$ was the first to be classified. This is the content of a classical result in \cite{lich} if $M$ is assumed to be diffeomorphic to ${\mathbb R}^3$ (the proof extends easily to all topologies). After many partial results it was recently shown by Masood-ul-Alam \cite{spherical} that when matter is composed of a perfect fluid we must have $n=1$ and the spacetime is spherically symmetric; in particular, Schwarzschild in the vacuum region. These spherical models have been studied extensively \cite{mark}. Solutions for $n=1$ without (spatial) symmetries, for sources composed of ideally elastic material have been constructed in \cite{abs}. For black holes it is known that $n$ has to be $1$ and the solution is isometric to the exterior of a Schwarzschild black hole. This has been shown in \cite{bunting} in the nondegenerate case and in \cite{chrusciel} generally.

\section{\bf{Separating surfaces}}
Let $(M,g)$ be an asymptotically flat Riemannian three manifold. We allow the possibility that $M$ has a finite number $q\geq 1$ of ends $M_\alpha$, $1\leq \alpha\leq q$, each being asymptotically flat. Recall that the static vacuum equations are given by $VR_{ij}-V_{ij}=0$ and $\Delta V=0$ for a positive function $V$ where $R_{ij}$ denotes the Ricci tensor of $g$ and $V_{ij}$ the covariant hessian of $V$ taken with respect to $g$.
We will be interested here in metrics which are static vacuum solutions outside a compact set, and at the very
least have nonnegative scalar curvature everywhere.

We will consider a surface $S$ which is noncompact, connected and properly embedded
in $M$. We first show that if such a surface is totally geodesic, then it has a finite number of ends
each of which is asymptotic
to a plane in one of the asymptotically flat ends of $M$ at infinity. Precisely we show that there is a compact subset $K$ of $M$ such that for each $\alpha$, $M_\alpha\cap(S\setminus K)$ is equal to a finite union of graphs of functions $f_p$, $1\leq p\leq k_\alpha$, over a Euclidean plane (in suitable coordinates
on $M_\alpha$) such that $f_p$ approaches a constant and its derivatives decay at an appropriate rate.

\begin{prop} Let $S$ be a noncompact, connected, totally geodesic surface properly embedded in
$M$. Assume that $S\cap M_\alpha$ is unbounded in the end $M_\alpha$. There exist asymptotically flat coordinates defined on $M_\alpha$ such that outside a compact set $K$ the
surface $S\cap (M_\alpha\setminus K)$ is the union of $k_\alpha\geq 1$ graphs of functions $x^3=f_p(x^1,x^2)$ for $1\leq p\leq k_\alpha$ such that  there are constants $a_p$ so that $f_p-a_p$ decays like $1/r'$ and the derivatives of the $f_p$ decay correspondingly faster, where $r'=r'_\alpha=\sqrt{(x^1)^2+(x^2)^2}$. Note that this description holds for each of the ends $M_\alpha$ for which $S\cap M_\alpha$ is unbounded and the number $k_\alpha$ depends on $\alpha$ as do the coordinates and the functions
$f_p$. (We take $k_\alpha=0$ if $S\cap M_\alpha$ is bounded.) We omit the dependence of the
coordinates and the $f_p$ on $\alpha$ for notational convenience.

Moreover, for $\sigma$ sufficiently large the compact subset of $S$ given by $S_\sigma=S\cap (K\cup
(\cup_{\alpha=1}^q\{r'_\alpha\leq \sigma\}))$ is a compact surface with boundary curve $C_\sigma$ (having $k=\sum_{\alpha=1}^q k_\alpha$ components) such that the Euler characteristic $\chi(S_\sigma)$ is equal to $\chi(S)$ and $\lim_{\sigma\to\infty}\int_{C_\sigma}\kappa\ ds=2\pi k$ where $\kappa$ is the geodesic curvature of the oriented curve $C_\sigma$ in $S$.
\end{prop}
\begin{pf} Our argument will work separately on each end, so throughout we focus attention on
one end $M_\alpha$ such that $S\cap M_\alpha$ is unbounded and we omit explicit reference to
$\alpha$ unless needed for clarity. From the work of \cite{beigstat} there exist coordinates on $M_\alpha$ defined outside a compact set $K$ such that $g$ is equal to a Schwarzschild metric up to order $r^{-2}$, that is
\[ g_{ij}=(1+2m/r)\delta_{ij}+O(r^{-2})
\]
where $m$ is the ADM mass. (We use the notation $O(r^{-k})$ to denote a term which is bounded
by a constant times $r^{-k}$ and whose derivatives up to a fixed order decay correspondingly faster.) Since $S$ is embedded and the manifold $M_\alpha\setminus K$ may be
chosen to be simply connected (for example we can take it to be diffeomorphic to the exterior of a ball in ${\mathbb R}^3$) it follows that $S$ is orientable. We choose the orientation for $M$ and hence for $S$
determined by the coordinates $x^1,x^2,x^3$, and let $e_1$ and $e_2$ be an oriented local orthonormal basis for $S$ relative to the metric $g$. It then follows that the length $N$ of
the $2$-vector $e_1\wedge e_2$ with respect to the Euclidean metric is $1+O(r^{-1})$. Therefore using the fact that $S$ is totally geodesic with respect to $g$ we have $D_{e_{\alpha}}[(e_1\wedge e_2)]=0$ for $\alpha=1,2$. Letting $\nabla$ denote the Euclidean connection, observe that the difference tensor $T=D-\nabla$ is of order $r^{-2}$ since it is given in Euclidean coordinates by the Christoffel symbols of
$g$, so we have
\[ 0=\nabla_{e_\alpha}(e_1\wedge e_2)+T_{e_\alpha}(e_1\wedge e_2).
\]
From this we see that $\nabla_{e_\alpha}(e_1\wedge e_2)$ is $O(r^{-2})$ and therefore
\[ \nabla_{e_\alpha}N=N^{-1}(\nabla_{e_\a}(e_1\wedge e_2))\cdot (e_1\wedge e_2)=O(r^{-2}).
\]
Now the length of the second fundamental form of $S$ with respect to the Euclidean metric is the Euclidean magnitude of $\nabla (N^{-1}e_1\wedge e_2)$ taken along $S$ (since $N^{-1}e_1\wedge e_2$ is the Euclidean unit tangent plane), and therefore the length of the Euclidean second fundamental form is $O(r^{-2})$.

{\bf Note:} The argument above shows that if $\hat{g}=\d+O(r^{-2})$,
then the magnitudes of the second fundamental form of $S$ taken with respect to the indicated metrics satisfy the inequality $|A_\d|\leq c|A_{\hat{g}}|+cr^{-3}$ since in this case the difference tensor is
$O(r^{-3})$.

Let $\sigma_0$ be a radius to be chosen large, and let $M_{\a,\s}$ denote the part of $M_\a$ {\it exterior} to
the open ball of radius $\s\geq\s_0$. Let $\e_0>0$ and consider the rescaled surface
$S(\s_0)=\e_0/\s_0(S\cap M_{\a,\s})\subset {\mathbb R}^3\setminus B_{\e_0}(0)$. The length of the second fundamental form of $S(\s_0)$ is then equal to $\s_0/\e_0$ times that of $S$ at corresponding points,
and distances are changed by a factor of $\e_0/\s_0$,
so we see that the second fundamental form of $S(\s_0)$ at a point $x$ is bounded by
$c(\e_0/\s_0)|x|^{-2}$. Since $S$ is connected, we see that either $S(\s_0)$ has a single component
without boundary or it has $k_\a\geq 1$ components $S_p(\s_0)$, $1\leq p\leq k_\a$, each with boundary
on $\p B_{\e_0}(0)$. In the former case it follows from Proposition 3.1 (next section) that for $\s_0$ sufficiently large (hence the second fundamental form small with quadratic decay), $S$ is the graph of a function $f$ over a plane which we
may take to be the $x^1x^2$-plane, and that the second derivatives of $f$ decay like $O((r')^{-2})$,
and the first derivatives like $O((r')^{-1})$. In the second case Proposition 3.1 implies that each of the $S_p(\s_0)$ may be
so described as the graph of a function $f_p$ with the same decay conditions. Note that since $S$
is embedded each of the $S_p(\s_0)$ is a graph over the {\it same} plane.

Scaling back to the original surface $S$ we obtain the description of $S\cap(M_\a\setminus K)$ as
a union of graphs. To get the required decay, we use the Schwarzschild form of the $1/r$ term
in the metric expansion. We observe that the metric $\hat{g}$ defined by $\hat{g}=(1+m/r)^{-2}g$ has the property that $\hat{g}=\d+O(r^{-2})$. Using the well known relation for second fundamental forms of
conformally related metrics we see
\[ A_g=A_{\hat{g}}+(1+m/r)^{-1}\hat{\nu}(1+m/r)\hat{g}
\]
where $\hat{\nu}$ denotes the unit normal of $S$ with respect to $\hat{g}$ and for a function $\varphi$,
we use $\hat{\nu}(\varphi)$ to denote the derivative of $\varphi$ in the direction $\hat{\nu}$.
Since $A_g=0$ and from the asymptotic behavior of the $f_p$ we see that
on the graph of $f_p$ we have $\hat{\nu}$ is plus or minus $\frac{\p}{\p x^3}+O(r^{-1})$, so we have
$|A_{\hat{g}}|=(\sqrt{3}m|x^3|/r^3)+O(r^{-3})$. From the fact that first derivatives
of $f$ decay like $O((r')^{-1})$ it follows that $f_p$ is bounded by $O(\log r')$. Putting $x^3=f_p$ in the
bound on the second fundamental form, we see that $|A_{\hat{g}}|=O((\log r)r^{-3})$. Since the metric
$\hat{g}$ is Euclidean up to terms of order $r^{-2}$, we use the Note above to improve the decay on the Euclidean second fundamental form to $O((\log r) r^{-3})$. This can then be used to show that $f_p$ is bounded and has a limit $a_p$ at infinity. Putting this information back into the second fundamental form bound tells us finally that the second derivatives of $f_p$ decay like $O((r')^{-3})$, and this implies the desired asymptotic decay.

The final statement on the behavior of the total geodesic curvature follows from the easily checked fact that the geodesic curvature of $C_\s$ is equal to $1/\s+O(\s^{-2})$ while the length of each component
of $C_\s$ is equal to $2\pi \s+O(1)$.
\end{pf}

\begin{thm} Assume that $M$ is static vacuum outside a compact set and has $R\geq 0$ everywhere.
Suppose there is a closed, noncompact, totally geodesic surface $S$ such that $g$ is static vacuum
in a neighborhood of $S$. It follows that $M$ is isometric to the Euclidean space ${\mathbb R}^3$.
\end{thm}
\begin{pf} Let $V$ be the static potential defined in a neighborhood of $S$ and outside a compact
set of $M$. We first show that $V$ is identically $1$ on $S$ and
that $S$ is flat (zero Gauss curvature). To see this, we choose a
local orthonormal frame so that the $e_\alpha$ are tangential for
$\alpha=1,2$ and $e_3$ is normal to $S$. We then take the tangential
trace of (\ref{Rij}) to obtain
\[ VR_{\alpha\alpha}=V_{\alpha\alpha}=\Delta_S V
\]
where we have used that fact that $S$ is totally geodesic to write the trace of the covariant derivatives
on $M$ in terms of the intrinsic Laplace operator on $S$. (It would be sufficient here that $S$ be minimal.)
Now the Gauss equation tells us that since $S$ is totally geodesic we have
\[ R_{\alpha\alpha}=R_{\alpha\beta\alpha\beta}+R_{\alpha 3\alpha 3}=2K+R_{33}
\]
where $K$ is the intrinsic Gauss curvature of the surface $S$. Since
$R=0$ in the static vacuum region due to (\ref{DelV}), this implies
that $R_{33}=-R_{\alpha\alpha}$, and therefore $R_{\alpha\alpha}=K$.
Thus we see that the restriction of $V$ to $S$ satisfies the
equation $\Delta_S V-KV=0$. Now we let $S_\sigma$ be as in
Proposition 2.1, and apply the Gauss-Bonnet theorem to obtain
\[ \int_{S_\sigma}K\ da=2\pi\chi(S)-\int_{C_\sigma}\kappa\ \ ds.
\]
The totally geodesic condition implies that $K=R_{1212}$ is bounded by a constant times $r^{-3}$,
and thus by Proposition 2.1, $K$ is an integrable function on $S$. Thus we may let
$\sigma$ tend to infinity to conclude $\int_S K\ da=2\pi\chi(S)-2\pi k\leq 0$ since $k\geq 1$ and the Euler characteristic of any connected noncompact surface is at most $1$. On the other hand we have
$K=V^{-1}\Delta_S V$, so we may also write
\[ \int_{S_p}K\ da=\int_{S_p}V^{-2}|\nabla_S V|^2\ da+\int_{C_p}V^{-1}\frac{\partial V}{\partial\nu}\ ds
\]
where $\nu$ is the outer unit normal along $C_p$. Since $V$ tends to $1$ and the derivatives of
$V$ decay at least as fast as $r^{-2}$ it follows that the boundary term goes to $0$ as $p$ goes to
infinity and we have
\[ \int_S K\ da=\int_S V^{-2}|\nabla_S V|^2\ da.
\]
We therefore conclude that the integral on the right is $0$ and hence $V$ is constant on $S$. It follows
that $V=1$ on $S$, and from the equation satisfied for $V$ that $K=0$ on $S$. It follows moreover that
$\chi(S)=1$, and hence $S$ is isometric to the Euclidean ${\mathbb R}^2$.

%
Now it is a known asymptotic property of the static equations (\cite{beig1978},\cite{beigstat}), that there is a constant $m$ so that
\[
V = 1 - \frac{m}{r} + o(\frac{1}{r^2})
\]
and that $m$ is equal to the ADM mass. Thus we have shown that $m$ is zero, so it follows from
the Positive Mass Theorem \cite{sy} that $M$ is isometric to the Euclidean ${\mathbb R}^3$. This
completes the proof.
\end{pf}
The following result is a consequence of Theorem 2.2.
\begin{thm} A nontrivial relativistic static $n$-body configuration cannot have a reflection symmetry
across a noncompact surface which is disjoint from the bodies.
\end{thm}
\begin{pf} Assume we had such a configuration with $S$ being the surface fixed by the symmetry $F$.
It would then follow that $S$ is totally geodesic since a geodesic $\sigma$ beginning at a point of $S$
and initially tangent to $S$ must remain in $S$ since $F\circ\sigma$ is a geodesic with the same initial
conditions and is therefore identical to $\sigma$. The result now follows from Theorem 2.2.

\end{pf}

\section{{\bf A technical result for surfaces in ${\mathbb R}^3$}}
In this section we prove the technical result used in the proof of Proposition  2.1. That result is the
following.
\begin{prop} Assume that $S$ is a closed, connected, noncompact, embedded surface in ${\mathbb R}^3\setminus B_{\e_0}$ where $B_r$ denotes the closed ball of radius $r$ centered at the origin.  Assume also that for any point $x\in S$ we have $|A|(x)\leq c\d_0 |x|^{-2}$ where $A$ denotes the second fundamental form of $S$. If $\e_0$ and $\d_0$ are sufficiently
small, then there exist Euclidean coordinates $x^1,x^2,x^3$ so that any connected component of
$S\cap ({\mathbb R}^3\setminus B_1)$ is contained in the graph of a function $x^3=f(x^1,x^2)$ defined for $r'=\sqrt{(x^1)^2+(x^2)^2}\geq 1/2$
such that the first and second derivatives of $f$ satisfy $|\partial f|\leq c(r')^{-1}$ and $|\partial^2 f|\leq c(r')^{-2}$.
\end{prop}
\begin{pf} We first consider the case in which $\overline{S}\cap\p B_{\e_0}=\emptyset$. In this case,
$S$ is a closed embedded surface in ${\mathbb R}^3$. Let $P\in S$ be a point nearest the
origin and note that $|P|>\e_0$. We choose Euclidean coordinates $y^1,y^2,y^3$ so that $P$
is at the origin and so that $\nu(P)=\frac{\p}{\p y^3}$ where $\nu$ denotes the unit normal vector field
to $S$. There is a neighborhood of
$0$ in $S$ which is the graph of a function $y^3=f_1(y^1,y^2)$ defined for $\rho'=\sqrt{(y^1)^2+(y^2)^2}\leq R$ so that $|\p f_1|\leq 1$. We show that the set of $R$ with this property consists of all positive real numbers, and thus the entire surface $S$ may be so represented. To see this, let $R$ be the largest radius for which such a representation is possible, and use the fundamental theorem of calculus
along the ray $\g(t)=(ty^1,ty^2,f_1(ty^1,ty^2))$ to write
\[ \nu(y^1,y^2,f_1(y^1,y^2))-\frac{\p}{\p y^3}=\int_0^1\frac{d}{dt}\nu(\g(t))\ dt.
\]
Since $|\p f_1|\leq 1$ it follows that $|\g'(t)|\leq \sqrt{2}\rho'$, and thus we have
\[ |\nu(y^1,y^2,f_1(y^1,y^2))-\frac{\p}{\p y^3}|\leq \sqrt{2}\rho'\int_0^1|A(ty^1,ty^2,f_1(ty^1,ty^2))|\ dt.
\]
Now $|ty^1,ty^2,f_1(ty^1,ty^2))|\geq t\rho'$, and thus from the second fundamental form bound we
have $|\nu(y^1,y^2,f_1(y^1,y^2))-\frac{\p}{\p y^3}|\leq c\d_0(\rho')^{-1}$. It follows that if $\d_0$ is
chosen sufficiently small we have $|\p f(y^1,y^2)|\leq 1/2$ for $\rho'\leq R$. This contradicts the
choice of $R$ as the largest radius for which $|\p f|\leq 1$. This shows that $S$ is globally given
as the graph of a function with gradient bounded by $1$. Therefore from the second fundamental
form bound we have $|\p^2 f_1|\leq c\d_0(\rho')^{-2}$. It follows by integration as above that
the first partials of $f_1$ converge to constants at infinity, and thus we may change coordinates
to $x^1,x^2,x^3$ so that $S$ is given as $x^3=f(x^1,x^2)$ and so that the first derivatives decay
like $(r')^{-1}$. This gives the desired conclusion under the assumption that $\overline{S}\cap\p B_{\e_0}=\emptyset$.

Let us now assume that $\overline{S}\cap\p B_{\e_0}\neq \emptyset$. We first analyze the points of $S$
which lie on the unit sphere. Let $P\in S\cap \p B_1$ and suppose that the tangent plane of $S$
at $P$ does {\it not} intersect $B_{2\e_0}$. If $\d_0$ is sufficiently small this implies that a  large neighborhood of $P$ on $S$ lies arbitrarily close to the tangent plane, and hence does not
intersect $B_{\e_0}$. In this case the argument above implies that a connected component of
$S$ is a global graph and hence we must have been in the first case. Therefore it follows that
the tangent plane to $S$ at $P$ intersects $B_{2\e_0}$, and therefore since $\e_0$ is arbitrarily small,
$\nu(P)$ is arbitrarily close to being tangent to the unit sphere. It follows from this that $S$ intersects $\p B_1$ transversally, and that the curves of intersection have small geodesic curvature. Since the curve of
intersection is embedded, we can see by elementary geometry that it must consist of a finite number of curves all of which lie in a small neighborhood of a great circle with each curve being $C^2$ close to the great circle.

Now if we consider a point $P$ on one of these curves $\g$, then we choose coordinates $y^1,y^2,y^3$
so that the point $P$ is $(1,0,0)$ and that $\nu(P)=\frac{\p}{\p y^3}$. A neighborhood of $P$ in $S$
may then be represented by the graph $y^3=f_1(y^1,y^2)$ with $f_1$ of small $C^2$ norm defined over a disk of radius $7/8$ centered at $(1,0)$. This representation then extends to cover a neighborhood of the curve $\g$ by the graph $y^3=f_1(y^1,y^2)$ defined for $1/4\leq \rho'\leq 3/2$. If we now consider the largest value of $R$ for which this representation extends to the set $1/4\leq\rho'\leq R$ with
$|\p f_1|\leq 1$, then we may repeat the argument above to show that $R=\infty$, and thus each of the intersection curves lies on a connected component of $S\cap ({\mathbb R}^3\setminus B_1)$ which has the required description as a graph of a function over the region $r' \geq 1/2$ in the plane. Note that the $1/4$ is replaced by $1/2$ since we need to do a slight rotation of coordinates to make the tangent plane at infinity to be the $x^1x^2$-plane. We could replace $1/2$ by any fixed small radius $r_0$
by taking $\e_0$ and $\d_0$ sufficiently small. Since $S$ is embedded, these planes must be parallel, so the description holds simultaneously for all components in a fixed system of Euclidean coordinates. This completes the proof.

\end{pf}

\end{document}